\documentclass[twocolumn]{article}
\usepackage{listings}
\usepackage{xcolor}
\usepackage{graphicx}
\usepackage{algorithm}
\usepackage[noend]{algpseudocode}
\usepackage{amsmath}
\usepackage{balance}
\usepackage{fancyhdr}
\usepackage{adjustbox}
\usepackage{url}

\usepackage[top=2cm, bottom=2cm, left=2cm, right=2cm]{geometry}

\lstdefinelanguage{SQL}{
    keywords={TYPE, DATASET, CREATE, FEED, WITH, START, STOP, TO, CHANNEL, BROKER, INDEX, FUNCTION, LET, GROUP, BY, SELECT, WHERE, FROM, ORDER, DESC, LIMIT, SUBSCRIBE, ON, AT, USING, AND, PERIOD, ACTIVE, PRIMARY, KEY, CONTINUOUS, CONNECT, PUSH, DURATION},
    keywordstyle=\color{blue}\bfseries,
    commentstyle=\color{gray},
    basicstyle=\ttfamily\small,
    showstringspaces=false
}

\title{\textbf{Optimizing Big Active Data Management Systems}}
\author{
    Shahrzad Haji Amin Shirazi\\
    \textit{University of California, Riverside}\\
    \texttt{Shaji013@ucr.edu}
    \and
    Xikui Wang\\
    \textit{University of California, Irvine}\\
    \texttt{xikuiw@uci.edu}
    \and
    Michael J. Carey\\
    \textit{University of California, Irvine}\\
    \texttt{mjcarey@ics.uci.edu}
    \and
    Vassilis J. Tsotras\\
    \textit{University of California, Riverside}\\
    \texttt{tsotras@cs.ucr.edu}
}
\date{} 

\begin{document}

\maketitle

\begin{abstract}
Within the dynamic world of Big Data, traditional systems typically operate in a passive mode, processing and responding to user queries by returning the requested data. However, this methodology falls short of meeting the evolving demands of users who not only wish to analyze data but also to receive proactive updates on topics of interest. To bridge this gap, Big Active Data (BAD) frameworks have been proposed to support extensive data subscriptions and analytics for millions of subscribers. As data volumes and the number of interested users continue to increase, the imperative to optimize BAD systems for enhanced scalability, performance, and efficiency becomes paramount. To this end, this paper introduces three main optimizations, namely: strategic aggregation, intelligent modifications to the query plan, and early result filtering, all aimed at reinforcing a BAD platform’s capability to actively manage and efficiently process soaring rates of incoming data and distribute notifications to larger numbers of subscribers.
\end{abstract}

\section{Introduction}
In today's fast-paced digital world, we are inundated with a tremendous amount of data every second. Managing and analyzing this ocean of data, widely known as Big Data, presents formidable challenges and numerous systems have been developed for addressing them. However, the majority of these systems operate in a \textit{passive} mode, merely processing and returning data in response to user queries.
This passive approach often falls short for users who not only want to analyze data but also \textbf{actively} receive updates on new data items that interest them, explore their \textbf{relationships} with other data, and even \textbf{enrich} them with additional information existing in different datasets.
These demands have led to the creation of Big Active Data (BAD) frameworks \cite{bad_DEBS, bad_to_the_bone, wang2020subscribing} that aim to support extensive data subscriptions and analytics for millions of subscribers.
A BAD framework circumvents the inefficiencies of cobbling together multiple independent systems (each dealing with a part of the needed processing, i.e., accessing Big Data, managing incoming streaming data, matching subscribers to information etc).

As we confront the relentless expansion of data volumes and the burgeoning number of users interested in this data, the challenge to manage ever-larger datasets becomes increasingly acute.  
This work focuses on optimizing a BAD platform
to better handle soaring rates of incoming data and a growing roster of subscriptions. By deploying a suite of optimization techniques including strategic aggregation, intelligent modifications to the query plan, and early result filtering  we have enhanced the BAD framework to its most optimized form yet. These improvements not only streamline data processing without adding more resources, but also significantly broaden the system's ability to serve a larger and more diverse subscriber base, marking a significant stride forward in the ongoing quest to \textit{activate} Big Data.
 \section{Related Work}
Tapestry \cite{tapestry} first introduced Continuous Queries as queries that are issued once and then return results continuously as they become available. Tapestry also defined continuous query semantics and created rewrite rules for transforming user-provided queries into incremental database queries. Subsequent research has primarily concentrated on queries involving streaming data. NiagaraCQ \cite{niagara_cq} enhanced the scalability and efficiency of continuous queries by breaking them down into smaller, manageable components and clustering similar queries based on their expression signatures. It organized signature constants in a specialized table and used joins to process similar queries as a group. Furthermore, to enhance computational efficiency, the system employed delta files that allow for incremental updates and evaluations of the data. STREAM represents a research prototype that was designed to handle continuous queries across both data streams and persistent storage \cite{Stream}. It provided a Continuous Query Language (CQL) for constructing continuous queries against streams and updatable relations \cite{TheCQLcontinuousquerylanguage}. Most continuous query projects have struggled with scalability, making them less suitable for Big Data use cases, as they often fail to scale effectively in distributed environments. 

Streaming engines can be used for data processing and data customizing pipelines and to provide real-time analytics. 
Apache Kafka \cite{kafka,kafka2020streams}, Apache NiFi \cite{apache_nifi}, Apache Flink \cite{flink}, and Amazon Kinesis \cite{amazonkinesis} are prominent platforms designed to handle and process large-scale, real-time data streams. 
Similarly, Azure Stream Analytics \cite{azurestreamanalytics} and Google Cloud Dataflow \cite{google_cloud_dataflow} are specialized for stream processing and real-time analytics.
These systems are optimized for high-throughput and low-latency processing, enabling them to handle vast amounts of data generated in real-time. However, they are not inherently equipped to provide long-term data storage solutions. Instead, these systems typically require integration with external storage solutions to persist processed data for later use or analysis.
This "glued-systems" approach has been shown to have performance disadvantages in \cite{wang2020subscribing}.

Traditional publication/subscription (pub/sub) systems \cite{pubsub, pub-sub-content, pub-sub-multijoin, pub-sub-profile-pruning, carzaniga2001, carzaniga2011} allow subscribers to register their interests in events and to be asynchronously notified about events from publishers. Although pub/sub services can handle a large number of subscribers, users often have to integrate such services with other systems for data processing, and complex computations across multiple data sources are not supported.

The systems mentioned above generally face challenges in scaling, persisting data, or handling complex subscription queries, which then requires integrating them with additional systems for effective data processing. The BAD platform, summarized in the next section, was designed and built to address these constraints. In this current work, we focus on further optimizing BAD. One of the key steps involves creating the BAD index, an idea that is conceptually similar to partial indexing \cite{PartialIndexing}; it focuses on indexing only the most frequently accessed or queried segments of a dataset, yet is distinct from partial indexing in its implementation as explained in Section~\ref{BADidx}.

\section{BAD Preliminaries}
\label{sec:what is BAD}
The BAD platform can enable millions of users to subscribe to data of interest and receive updates continuously. It is different than continuous queries, streaming engines and pub/sub systems as it also supports Big Data analytics with a declarative language, SQL++ (a SQL-inspired query language for semi-structured data \cite{sqlpp2024,fang2023}).

\begin{figure}[!ht]
    \centering
    \includegraphics[width=1\columnwidth]{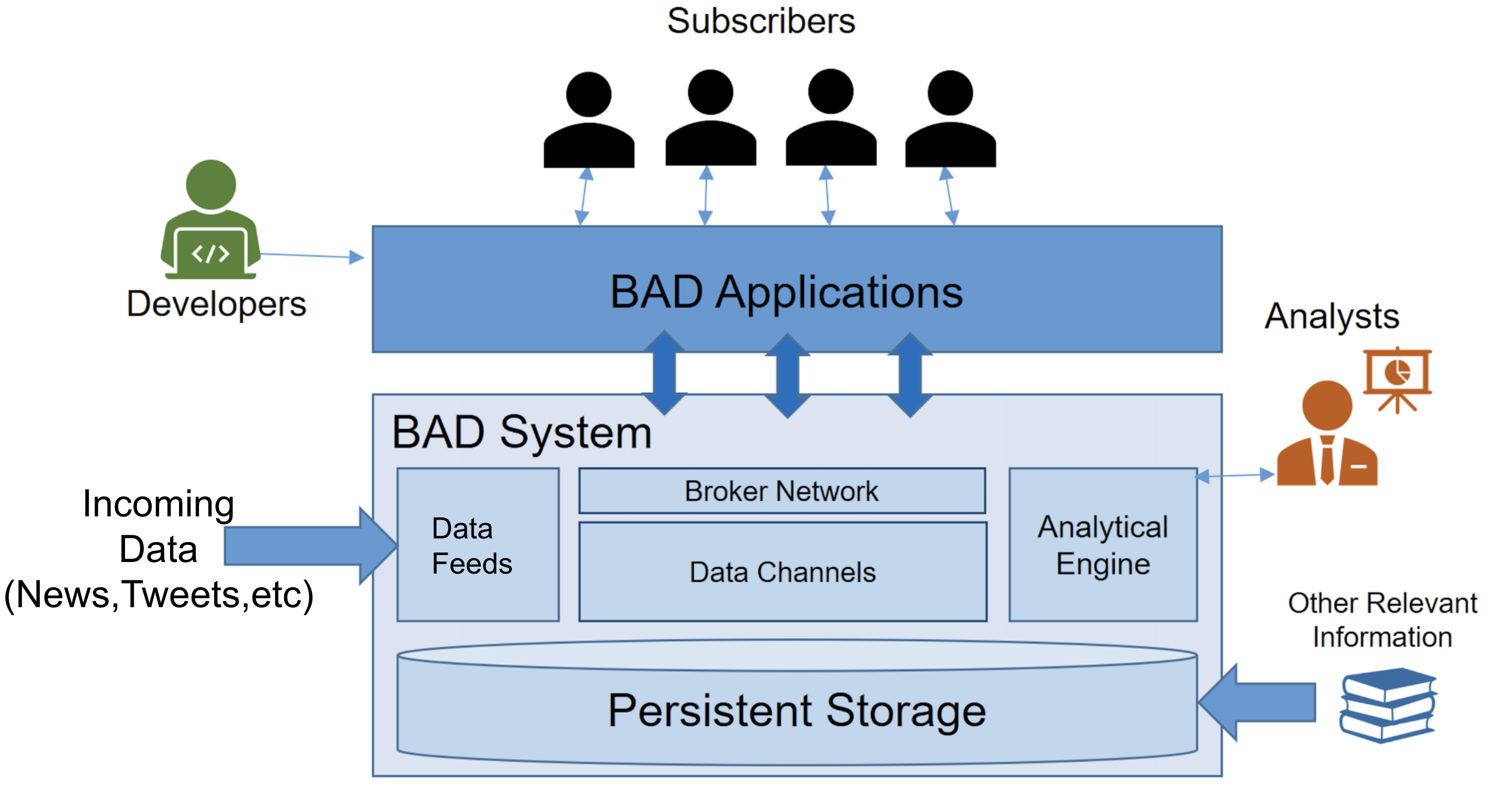}
    \caption{An overview of the BAD platform.}
    \label{fig: An overview of the BAD system}
\end{figure}

An overview of the BAD platform is shown in Figure \ref{fig: An overview of the BAD system}. 
BAD has five basic blocks \cite{jacobs2017bad}: (i) Data Feeds, which manage the ingestion of rapidly arriving data; (ii) Persistent Storage, responsible for storing the data; and an (iii) Analytical Engine, which enables analytic queries on the stored and incoming data to reveal useful information. Additionally, BAD includes (iv) \textit{Data Channels} and (v) \textit{Brokers}, which are important for this paper and are described in detail later in this section using an example application.

There are 3 different types of users for the BAD system; \textit{Subscribers} who subscribe to channels so they can get the data of interest, \textit{Developers} who create the BAD channels, and \textit{Analysts} who run queries on the data stored in the system.

 For this work we obtained a copy of the BAD open source platform and have extended it further. The existing implementation of the BAD platform \cite{bad_to_the_bone,bad_application, wang2020subscribing} was built as an extension of Apache AsterixDB \cite{asterix}, a Big Data Management System (BDMS) that provides distributed data management for large-scale, semi-structured data.

In AsterixDB, data is stored by creating a Datatype, which describes known aspects of the data being stored, and then a Dataset, which is a collection of records with a given datatype. As an example, Figure \ref{ddl:creating datatype} shows the DDLs that create a datatype named {\tt \small "EnrichedTweet"} and a dataset named {\tt \small "EnrichedTweets"} based on that type.
As the name implies, this dataset includes enriched tweets that are derived from original tweets \cite{alkowaileet2018enhancing,grover2015data,wang2019idea}, augmented with additional fields that contain specific information extrapolated from the text of the tweets (like threatening\_rate, weapon\_mentioned etc.)
The keyword "ACTIVE" used in the DDL for creating the dataset {\tt \small EnrichedTweets} enables continuous query semantics.

 \begin{figure}[!ht]
\footnotesize
\begin{lstlisting}[
           language=SQL,
           basicstyle=\ttfamily,
           showstringspaces=false,
           commentstyle=\color{gray}
        ]
CREATE TYPE EnrichedTweet AS {
 tid:int,
 text:string,
 retweet_count:int,
 threatening_rate:int,
 hate_speech_rate:int,
 retweeted_status:string,
 weapon_mentioned:boolean,
 drug_activity:string,
 about_country:string,
 state:string,
 location:point,
 additional_info:string};
CREATE ACTIVE DATASET 
EnrichedTweets(EnrichedTweet) PRIMARY KEY tid;
\end{lstlisting}
\caption{DDLs for creating a Datatype and its Dataset.}
\label{ddl:creating datatype}
\end{figure}

\subsection{A BAD Application Example}
Tweets can offer valuable insights into public opinions and behaviors, with many people relying on them to stay informed about important topics. However, given the volume and velocity of tweets generated every second, it can be very challenging to extract useful information from them. Therefore, providing a way for users to focus on tweets that are specifically relevant to their interests is highly beneficial.
People may be particularly interested in tweets related to specific topics such as sports, politics, or various types of crime. The BAD platform addresses this need by creating channels that users can subscribe to, delivering relevant data based on their preferences, and leveraging a broker network to ensure that tweets are delivered to subscribers in real time. This system enables different applications to effectively manage the vast flow of information, allowing users to focus on the content that matters most to them.
\subsection{Brokers}
To facilitate the distribution of results to millions of subscribers, the BAD platform integrates a broker sub-system to handle both subscription communication and
result delivery \cite{bad_to_the_bone}. Brokers can range from individual servers dedicated to relaying custom data to subscribers, to complex networks offering features like load balancing, subscription handover, and varied caching methods. Different brokers can be registered as HTTP endpoints in the BAD platform, and subscribing end users have the flexibility to select a broker that aligns with their specific requirements. Once results are prepared for subscribers, the BAD platform efficiently dispatches the relevant updates to all subscribed individuals (end users) through the designated brokers.
\subsection{Data Channels}
The BAD platform user model exploits the shared structure among subscriptions and offers it as a service, namely as a \textit{data channel}. Data channels allow developers to activate parameterized queries as services for users to subscribe to and continuously receive their data of interest. In practice, the need for similar information among users would likely result in the creation of comparable queries. 

Consider a scenario where users might want the system to "send them threatening tweets that relate to the US, are widely retweeted, and the sender's location is close to their location".
The {\tt \small TweetsAboutCrime} channel, depicted in Figure \ref{DDL: creating chanel3}, allows its subscribers to receive nearby enriched tweets that concern the United States, possess a retweet\_count greater than 10,000, and have a threatening\_rate exceeding 5 (on a scale of 0 to 10). The term \textbf{PERIOD} refers to how often the channel is executed, which will be described later.
To ensure that channels only process and deliver newly incoming tweets between executions, the \textbf{is\_new} function is employed in the channel definition to exclude already processed tweets and thus implement continuous query semantics. The channel’s parameters, such as MyUserName in the {\tt \small TweetsAboutCrime} channel, allow the system to personalize results for each subscriber. Users can subscribe to channels of interest by providing parameters through DDL commands, as demonstrated in Figure \ref{DDL: subscribing to a channel}. In this example, the user selects "user123" as the parameter. The system retrieves the subscriber's location from the {\tt \small UserLocations} dataset by matching the username and will provide results for the user based on their location.

BAD data channels provide two modes for delivering data: \textit{push} and \textit{pull}. In push mode, the data of interest is pushed to brokers directly. In  pull mode, the broker will
receive a notification from the channel when there exists new data of interest for its subscribers; the subscribers can then ask for (pulling) their data at any given time. This paper focuses exclusively on push channels, that is, the broker receives the produced data immediately after each channel execution and disseminates (pushes) the data to subscribers. We defer the exploration of optimizing pull channels to future work.

\begin{figure}[!ht]
\footnotesize
\begin{lstlisting}[
           language=SQL,
           basicstyle=\ttfamily,
           showstringspaces=false,
           commentstyle=\color{gray}
        ]
// Users can subscribe to this channel using 
// their usernames to get the threatening tweets  
// posted near their location.       
CREATE CONTINUOUS PUSH CHANNEL  
TweetsAboutCrime(MyUserName)
PERIOD duration ("PT10M") {
 SELECT t.text
 FROM EnrichedTweets t, UserLocations u 
 WHERE spatial_distance(u.location,t.location)<10
       AND u.username=MyUserName
       AND t.about_country="US"
       AND t.retweet_count>10000
       AND t.threatening_rate>5 
       AND is_new(t)};
\end{lstlisting}
\caption{DDL for the {\tt \small TweetsAboutCrime} channel.}
\label{DDL: creating chanel3}
\end{figure}

\begin{figure}[!ht]
\footnotesize
\begin{lstlisting}[
           language=SQL,
           basicstyle=\ttfamily,
           showstringspaces=false,
           commentstyle=\color{gray}
        ]
SUBSCRIBE TO 
TweetsAboutCrime("user123") ON BrokerA;
\end{lstlisting}
\caption{DDL for subscribing to the channel {\tt \small TweetsAboutCrime}. }
\label{DDL: subscribing to a channel}
\end{figure}
\subsubsection{Channels Under the Hood}
This section sets the stage for our work by detailing the BAD platform's infrastructure for creating and managing data channels, highlighting the {\tt \small TweetsAboutCrime} channel. 
Considering the need to promptly inform subscribers, the update interval for the {\tt \small TweetsAboutCrime} channel ("period" in Figure \ref{DDL: creating chanel3}) has been set to every 10 minutes. Upon channel creation, a dataset named {\tt \small TweetsAboutCrimeSubscriptions} is created for storing subscriptions. Each subscription is identified by a unique ID with an associated broker and subscription parameters. Every 10 minutes, a recurring query, as shown in Figure \ref{query: under the hood of a channel}, generates channel results and matches them with the relevant subscriptions for distribution via brokers. The broker information is stored in the  {\tt \small Broker} metadata dataset.
 \begin{figure}[!ht]
\footnotesize
\begin{lstlisting}[
           language=SQL,
           basicstyle=\ttfamily,
           showstringspaces=false,
           commentstyle=\color{gray}
        ]
SELECT result, current_datetime()
       as deliveryTime, sub.subscriptionId as sId
FROM Metadata.`Broker` b, 
     TweetsAboutCrimeSubscriptions sub,
     TweetsAboutCrime(sub.param0) result
WHERE result.BrokerName=b.BrokerName; 

\end{lstlisting}
\caption{The query running under the hood for the channel {\tt \small TweetsAboutCrime}.}
\label{query: under the hood of a channel}
\end{figure}
\section{Optimizing BAD}
To address the challenges posed by escalating data volumes and a rising count of subscriptions, it is imperative to implement optimization strategies for the channels within the BAD platform. These optimizations are crucial for ensuring that the system can manage data effectively, accommodate all subscriber requests promptly, expedite query processing, and consistently meet designated delivery deadlines as its users and data scale.

When examining the processing performed by the original implementation of the BAD platform, we discovered three patterns which offer opportunities for optimization. In particular,
\textbf{(1) Duplicate Processing}: the BAD platform processes results for subscriptions that ask for identical parameters as if they were distinct, resulting in redundant computations. \textbf{(2) Overprocessing}: subscription queries are executed on the entire dataset that has accumulated since the last execution, even when new records do not match existing subscriptions, leading to superfluous processing. \textbf{(3) Late Data Filtering}: In the current BAD platform, although the query and all predicates of a channel are defined during channel creation, which occurs prior to any channel execution, the system postpones processing and identifying the relevant data until execution time. For each of these cases, we discuss the proposed solutions and their benefits below, which are then showcased in the experimental section.
\subsection{Duplicate Processing: Aggregating Subscriptions}
In systems designed to serve a large user base, multiple users often create similar queries, leading to redundant processing when retrieving results separately for each user. The BAD platform introduced data channels to group common user query patterns into a single parameterized query, allowing users to select their own parameters. However, further analysis has revealed additional areas where we can go a step further by implementing even more efficient sharing mechanisms.
For instance, in use cases where users subscribe to specific content categories like trending topics or news alerts, the limited number of categories can result in numerous subscriptions requesting the same data, differentiated only by subscription IDs. This redundancy can burden the system with multiple subscriptions that, upon closer inspection, are requesting the same information. 
A similar issue is observed in the BAD platform. For example, in the {\tt \small TweetsAboutDrugs} channel, shown in Figure \ref{DDL: creating chanel2}, subscribers must specify their state. Since there are only a finite number of U.S. states, this setup often results in multiple records for the same state, with the only variation being the users IDs.
\begin{figure}[!ht]
\footnotesize
\begin{lstlisting}[
           language=SQL,
           basicstyle=\ttfamily,
           showstringspaces=false,
           commentstyle=\color{gray}
        ]
CREATE CONTINUOUS PUSH CHANNEL  
TweetsAboutDrugs(Mystate)
PERIOD duration ("PT10M") { 
 SELECT t.text
 FROM EnrichedTweets t 
 WHERE t.state=Mystate 
       AND t.threatening_rate=10
       AND t.drug_activity="Manufacturing Drugs" 
       AND is_new(t)};
\end{lstlisting}
\caption{DDL for the {\tt \small TweetsAboutDrugs} channel.}
\label{DDL: creating chanel2}
\end{figure}

To avoid the inefficiencies associated with storing duplicate records, we propose grouping subscriptions based on their parameters and associated brokers. As a result, we create subscription-group records that maintain the group's parameter and broker name, along with an array that includes all subscription IDs requesting the respective parameter and broker. Since groups may vary in size, these records are of variable length.

\begin{figure}[!ht]
  \centering
  \includegraphics[width=1\columnwidth, height=0.35\textheight]{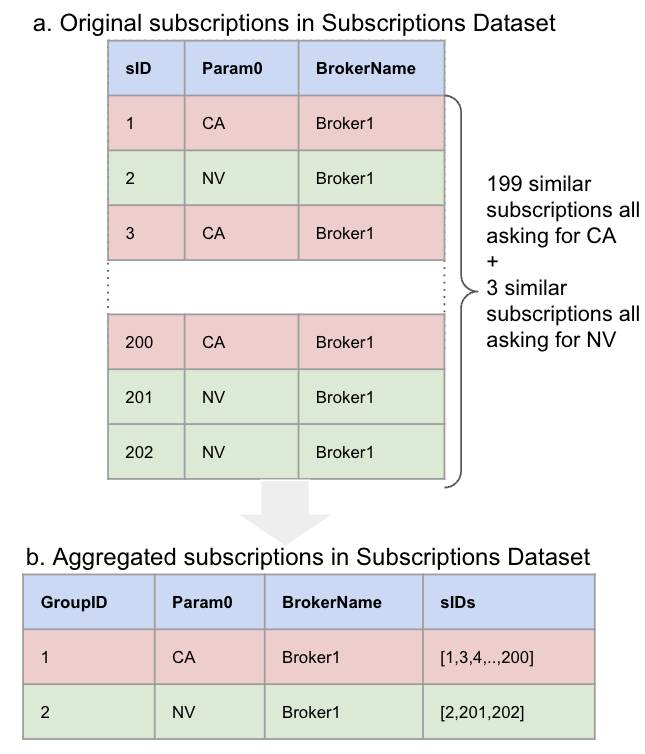}
  \vspace{-\baselineskip}
  \caption{Aggregation of subscriptions based on matching parameters and brokers.}
  \label{fig:aggregating_subscription}
\end{figure}

Figure~\ref{fig:aggregating_subscription}(a) illustrates the original subscription dataset, while Figure~\ref{fig:aggregating_subscription}(b) depicts how it is transformed for the new optimized system. To minimize additional overhead when grouping subscriptions, the groups are created, and subscriptions are assigned to the appropriate group as they enter the system. The ID of each new incoming subscription is either allocated to a pre-existing group, or it initiates a new subscription-group if its parameter and broker combination is not yet represented. 

Although grouping many subscriptions into a single record may seem practical, it introduces certain system challenges, which are discussed below along with potential solutions. Note that subscription aggregation can be applied in various domains where multiple users share common interests or query parameters. For instance, in a financial monitoring system, users may subscribe to get updates about stock market trends based on specific conditions such as price movements, trading volume, or company performance metrics. Grouping these subscriptions by similar thresholds or parameters allows the system to handle multiple users with similar interests more efficiently, reducing duplicate processing. 
Similarly, in a content recommendation system, users may subscribe to updates about certain genres, authors, or topics. Aggregating these subscriptions based on similar preferences ensures more efficient content delivery and query execution.
\subsubsection{Parallelism in AsterixDB}
\label{Improving Parallelism}
In all scalable database management systems, when data is overly aggregated, the ability to leverage the system’s parallel architecture can be compromised, as fewer tasks can be distributed simultaneously, leading to potential bottlenecks and decreased performance efficiency. AsterixDB faces similar challenges. 
To delve deeper into the operational mechanics, we first note that the unit of data which is consumed and produced by different tasks in AsterixDB is called a \textit{frame}. It is a fixed-size chunk of contiguous bytes which always contains complete records, ensuring that a record is not split across multiple frames. An operator that produces data packs a frame with a
sequence of complete records and sends it to the consumer operator who then interprets the records.

Overall, the frame size in AsterixDB is selected to balance memory efficiency, data movement, and task execution performance within the system. The optimal frame size is typically chosen based on the characteristics of the data being processed, the available memory resources, and the specific requirements of the workload. Larger frame sizes are often preferred for workloads involving complex data processing or large records, as they can reduce the number of I/O operations and enhance network efficiency. However, this choice must be carefully managed to avoid memory contention or excessive garbage collection, which can degrade system performance.

After fixing the frame size, if a record is larger in size than the standard frame size, the particular frame is enlarged to include this record; as a result, there may be frames that are longer than the fixed size.
Consolidating many subscriptions into a single, large record affects the distribution of processing tasks across different operators. 
Let \textit{f} denote the fixed frame size in bytes and \textit{s\_{i}} the record size in bytes of subscription-group \textit{i} .  
If the group consolidates many subscriptions and \textit{s\_{i}} surpasses \textit{f}, we must expand the frame size to fully encompass the subscription-group since a record cannot be fragmented across frames. This may lead to fewer but larger frames and potentially reduce parallelism.
In this case dividing subscription-group \textit{i} into smaller subgroups can be another option to improve parallelism.
However, dividing subscription-group \textit{i} into too many smaller subgroups will lead to significantly increasing the computational load, since the system will calculate the same result multiple times. 

Clearly, there is a trade-off involved in optimizing the number of subscriptions within each group. This trade-off will be further examined in the experimental section. Algorithm \ref{algorithm1} demonstrates the process of adding new subscriptions to the dataset after determining the optimal subgroup size. 
Subscriptions are assigned to groups dynamically as they enter the system. If no existing group matches the parameters and broker of interest, or if the matching group has reached capacity, the number of subscriptions allowed in each group is determined based on the frame size, and a new subscription group is created if required. Since subscriptions are added individually as users subscribe to a channel, this grouping process incurs minimal overhead. Furthermore, all grouping is completed before the execution of the next channel begins, ensuring runtime efficiency.

\begin{algorithm}
\caption{Handling a New Subscription}
\label{algorithm1}
\raggedright
\begin{algorithmic}[1]
\State \textbf{Input Variables:}
\State $sub$:the new subscription
\State $subGroup$:identified by $params$,$subIDs$ and $broker$
\State $subGroups$:the array of subGroups for the channel
\State $subIDs$:the array of subscription IDs in a 
subGroup

\State $broker$:broker name for a subGroup

\State $params$:array of parameters for a subGroup

\State $AcceptableGroupSize$:calculated based on frame size $F$

\vspace{0.25cm}

\State \textbf{Algorithm:}
\If{\textbf{not} \Call{AddToExistingGroup} {$subGroups$, $sub$}}
    \State $subGroups$.\text{addsubGroup}($sub.params, sub.ID$
        \Statex \quad $, sub.broker$)

\EndIf

\vspace{0.1cm}

\Function{AddToExistingGroup}{$subGroups$, $sub$}

    \State $groups \gets subGroups.get($sub.params$, $sub.broker$)$
\Comment{retrieves groups with same parameters and broker}
    \vspace{0.1cm}

    \For{$g$ in $groups$}
    
        \If{$g.subIDs.\text{Size()} \leq AcceptableGroupSize$}
            \State $g.\text{add}(sub.ID)$
            \State \Return \textbf{True}
        \EndIf
        
    \EndFor
    \State \Return \textbf{False}

\EndFunction

\end{algorithmic}
\end{algorithm}
\vspace{-20pt}

\subsubsection{Broker Benefits}
Aggregating subscriptions extends advantages beyond merely improving query execution times (the query execution time is defined as the duration measured from the start to the end of the query execution).  
It also offers significant advantages on the broker side by notably decreasing both the communication time between the BAD platform and the brokers, as well as the processing time required by brokers to manage the results prior to dispatching them to subscribers. Consider a scenario where a new {\tt \small EnrichedTweets} instance (which is around 32 KB) pertains to drug-related activities in California, and suppose there are one million subscriptions for this state on the {\tt \small TweetsAboutDrugs} channel. Previously, this would create 1 million individual but similar results, each corresponding to a subscription, thereby burdening the broker with the management of redundant outcomes. 

By aggregating the subscriptions according to channel parameters we substantially decrease the volume of results that need to be transmitted and processed by the broker. 
Instead of sending results for each individual subscription, results are sent out per group, decreasing the data volume from 32 GB to just 0.07756 GB.

\subsection{Overprocessing: Augmenting the Query Plan to Align with User Preferences}
\begin{figure}[!ht]
    \centering
    \includegraphics[width=1\columnwidth]{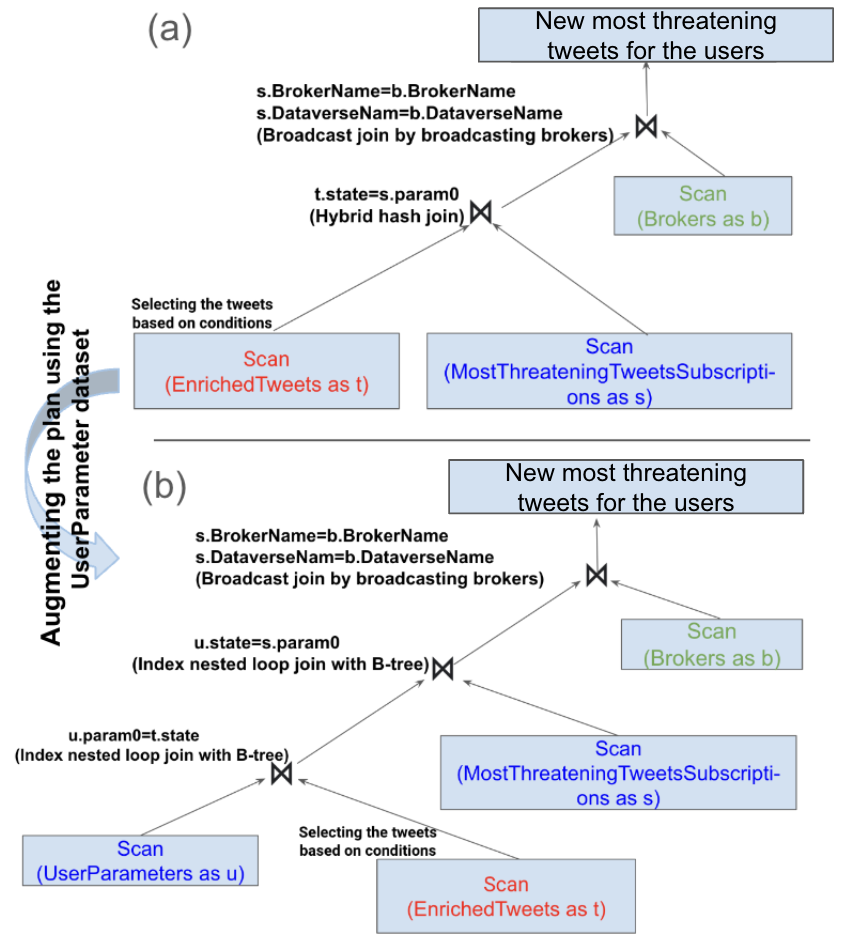}
    \caption{(a) Original vs. (b) Optimized channel plan.}
    \label{fig: query plan}
\end{figure}

Consider the {\tt \small MostThreateningTweets} channel, shown in Figure \ref{DDL: creating chanel}, which allows users to learn about the most threatening (level 10) tweets in their state.
The query plan for implementing this channel appears in Figure \ref{fig: query plan} (a). 
Originally, the BAD platform would completely scan the {\tt \small EnrichedTweets} dataset, the channel's subscription dataset called { \tt \small  MostThreateningTweetsSubscriptions}, and the {\tt \small Brokers} datasets, and eventually perform several steps leading up to a join between the selected {\tt \small EnrichedTweets} and the {\tt \small MostThreateningTweetsSubscriptions} dataset. 
This approach becomes wasteful if most of the incoming data rarely meets the subscription criteria. 
 \begin{figure}[!ht]
\footnotesize
\begin{lstlisting}[
           language=SQL,
           basicstyle=\ttfamily,
           showstringspaces=false,
           commentstyle=\color{gray}
        ]
CREATE CONTINUOUS PUSH CHANNEL 
MostThreateningTweets(MyState)
PERIOD duration ("PT10M") { 
 SELECT t.text
 FROM  EnrichedTweets t
 WHERE t.state=MyState
       AND t.threatening_rate=10 
       AND is_new(t)};
\end{lstlisting}
\caption{DDL of a channel created for the most threatening tweets.}
\label{DDL: creating chanel}
\end{figure}

This is the case when only a tiny fraction of tweets pertain to criminal activities and require law enforcement's attention. Therefore, failing to consider subscription parameters early in the query execution process can lead to creating a large volume of results which will eventually be discarded due to not matching any subscription criteria.
Additionally, it is important to note that some subscriptions might have similar parameters, and thus it is also crucial to structure the query process to avoid redundant computations.

This scenario can arise in any system that tries to align users' interests with incoming or stored data. Overloading operators with irrelevant data that doesn't contribute to the desired results is inefficient and undesirable.
We propose an approach that selects only the data satisfying at least one subscription. An initial solution might involve identifying relevant results by performing a join between the dataset containing the required parameters, e.g., the Enriched tweets dataset, and the subscription dataset in the first step. However this would result in a massive join operation between two datasets with millions of records but with a small number of results. For instance, in the case of the {\tt \small MostThreateningTweets} example, this solution would involve joining {\tt \small EnrichedTweets} with the subscriptions, leading to a large join that produces a small number of results.

To address effectively, we introduce a {\tt \small UserParameters}  dataset (a dataset which will be created by the system when a channel is created), integrated into each channel's plan, replacing the direct use of subscription parameters. This dataset includes fields for the channel's parameter(s) and the number of subscriptions interested in each. These fields facilitate the dynamic addition or removal of parameters as subscriber interests evolve.


The updated query advances the join operation between the {\tt \small UserParameters} and {\tt \small EnrichedTweets} datasets to occur during the initial data scan, reducing processing overhead. To further enrich channel results with detailed subscription information, such as each subscriber's broker, an index nested loop join is applied between the outcomes derived from the above join and the {\tt \small MostThreateningTweetsSubscriptions} dataset. The revised query plan is depicted in Fig \ref{fig: query plan} (b).



\subsection{Late Data Filtering: BAD Index}
When a query is known in the system before execution, it creates the possibility for the system to filter out irrelevant incoming records based on the query’s specific conditions, potentially improving execution efficiency. We can observe a similar scenario in the BAD system, where channels incorporate fixed selection criteria to ensure only relevant data is processed and delivered to subscribers. For example, 
the {\tt \small TweetsAboutCrime} channel appearing in Figure \ref{DDL: creating chanel3} has three predicates that involve the incoming data (combined with AND): {\tt \small t.about\_country="US",} {\tt \small t.retweet\_count>10000,} {\tt \small t.threatening\_rate>5}.  
Knowing these fixed predicates in advance allows for the integration of an additional pre-execution filter, streamlining the query evaluation process by focusing only on relevant data from the outset before executing the main logic of the query.
This filter identifies an incoming record that meets all fixed selection predicates specified in the channel query's WHERE clause and adds this record's id in 
a dedicated secondary index, known as the channel's \textbf{BAD index}.
The method for creating the BAD index, along with the changes required to adapt the query plan to utilize it, are described below.

\subsubsection{Creating the BAD index}
\label{BADidx}
When a channel query is submitted to the system, a BAD index is created for each active dataset involved in the channel that has fixed selection predicates applied to it. The fixed conditions for each dataset are grouped together and added to an ordered list, called {\tt \small conditionsList}, created for that active dataset, and when the channel is deleted from the system its conditions are also deleted. Conditions lists help determining whether incoming data satisfies the conditions for any channel. Each new incoming record will be checked against all groups of conditions in this list, and if it satisfies all conditions for a given channel, it will be added to that channel’s BAD index.
For example, when the {\tt \small TweetsAboutCrime} channel is created (Figure \ref{DDL: creating chanel3}), it includes three fixed conditions on the active {\tt \small EnrichedTweets} dataset. Consequently, the channel’s conditions are added to the {\tt \small EnrichedTweets} dataset’s {\tt \small conditionsList} and a BAD index called {\tt \small TweetsAboutCrimeBADindex} is created for the channel. Any new tweet that meets all three predicates in the {\tt \small TweetsAboutCrime} channel will be added to the {\tt \small TweetsAboutCrimeBADindex}. 

\begin{figure*}[!ht]
    \centering
    \includegraphics[width=\textwidth, height=0.26\textheight]{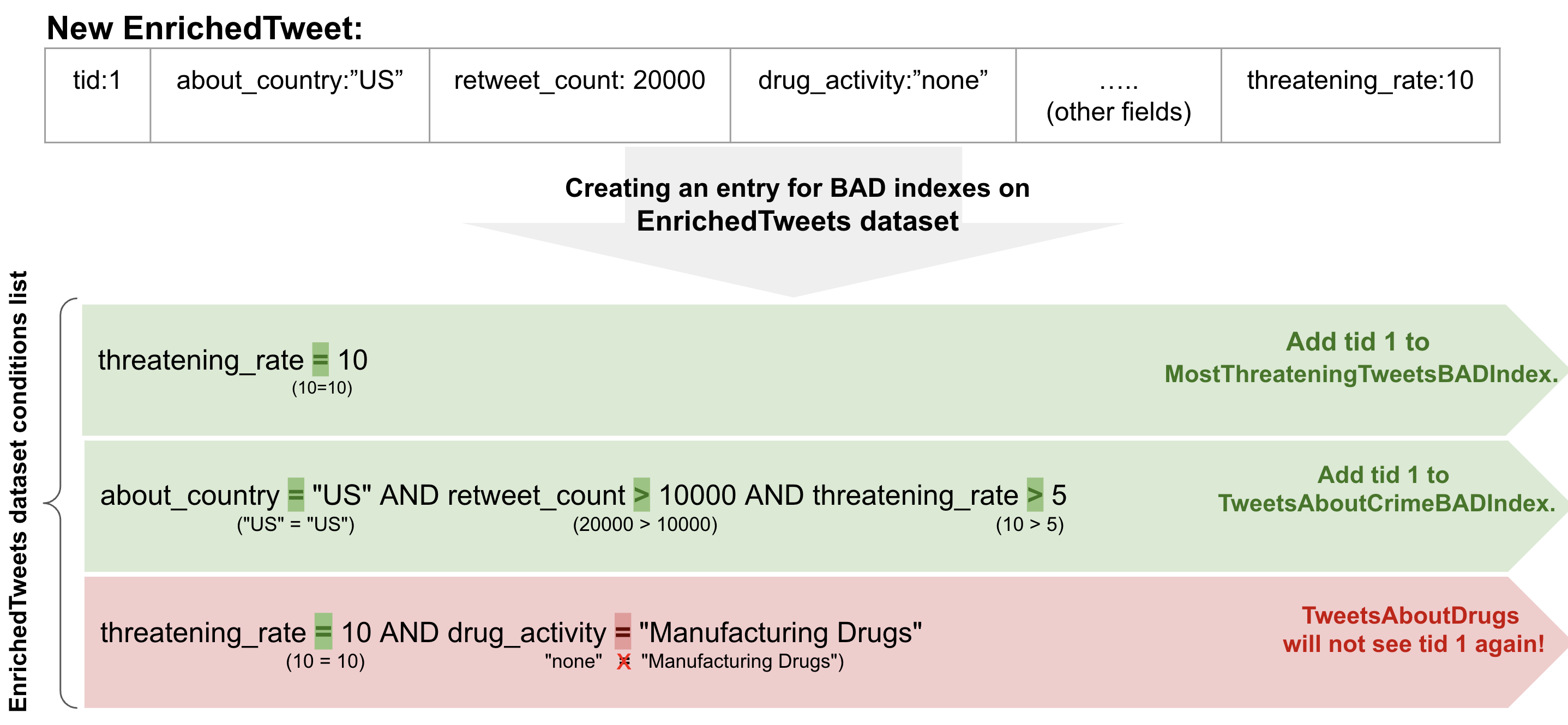}
    \caption{Processing a new incoming tweet and inserting it into the BAD indexes for each channel }
    \label{fig: adding new data to BADindex}
\end{figure*}

Algorithm \ref{algorithm2} demonstrates the procedure for adding entries to BAD indexes for an active dataset. Figure \ref{fig: adding new data to BADindex} illustrates how a tweet is processed and integrated into the BAD indexes, with all the channels created in the previous sections present in the system.

Unlike general-purpose indexes, which store data for all records, the BAD index focuses solely on records that meet the fixed predicates of a channel's query. 
The BAD index offers a major improvement over traditional indexes by filtering and retrieving only the data relevant to a specific query (we call this \textit{`early result filtering}'), thus avoiding the inefficiencies of indexing all records in a dataset. This selective indexing reduces storage overhead and eliminates the need to scan irrelevant data, resulting in faster query execution. 
This indexing approach facilitates the application of time filters \cite{timeFilter} to the BAD indexes, enabling efficient retrieval of only the most recent tweets when the is\_new function is present the query. The time filter employs the timestamp from the channel's most recent execution to guarantee that the index search includes only records that have arrived at or since the time of the last execution. As a result, a BAD index boosts channel execution by quickly locating data that meets the channel criteria, eliminating the need to scan the entire dataset ({\tt \small EnrichedTweets} in the above example). 

Note that this pre-processing is reminiscent of, but different from, 
Partial Indexing \cite{PartialIndexing}, which focuses on indexing only the most
frequently accessed or queried segments of a dataset. A BAD index does not index any attributes of the dataset; instead it consolidates the the primary keys of all records that satisfy all the fixed predicates of a channel.
This enables quicker access to the relevant records as opposed to scanning the entire dataset or using a traditional index which includes entries for all records, not just the ones that meet the specific criteria of the channel.
\begin{algorithm}[!ht]
\caption{BAD Index Record Insertion}
\label{algorithm2}
\raggedright
\begin{algorithmic}[1]
\State \textbf{Input Variables:}
\State $rec$ : the incoming  record
\State $ds \gets rec.getDataset()$ 
\Comment{Dataset in which the record is being inserted}
\State $conds \gets ds.getConditionsList()$ 
\Comment{List of fixed conditions for each channel}
\State $channelNames \gets ds.getChannelNames()$
\Comment{Names of channels with fixed predicates on the dataset}
\State $indexList \gets ds.getBADIndexList()$
\Comment{List of BAD indexes for the dataset}

\State \textbf{Algorithm:}
\For{each $i \gets 1$ \textbf{to} $conds.size()$}  
    \Comment{Check if the record satisfies all conditions for channel $i$}
       \vspace{0.05cm}

    \If{\textsc{checkConditions}($rec$, $conds[i]$)}
        \State $index \gets indexList.get(channelNames[i])$
        \State $index.addRecord(rec)$ 
        \Comment{Add the record to the corresponding index} 
    \EndIf
    \vspace{0.1cm}

\EndFor
\vspace{0.1cm}
\Function{checkConditions}{$rec$, $condGrp$}
    \For{each $j \gets 1$ \textbf{to} $condGrp.size()$}  
        \If{ \textbf{not} \textsc{satisfy}($rec$, $condGrp[j]$)}
            \State \Return \textbf{False}
        \EndIf
    \EndFor
    \State \Return \textbf{True}
\EndFunction

\end{algorithmic}
\end{algorithm}
\subsubsection{ Changing the Query Plan}
\begin{figure*}[!ht]
    \centering
    \includegraphics[width=\textwidth, height=0.28\textheight]{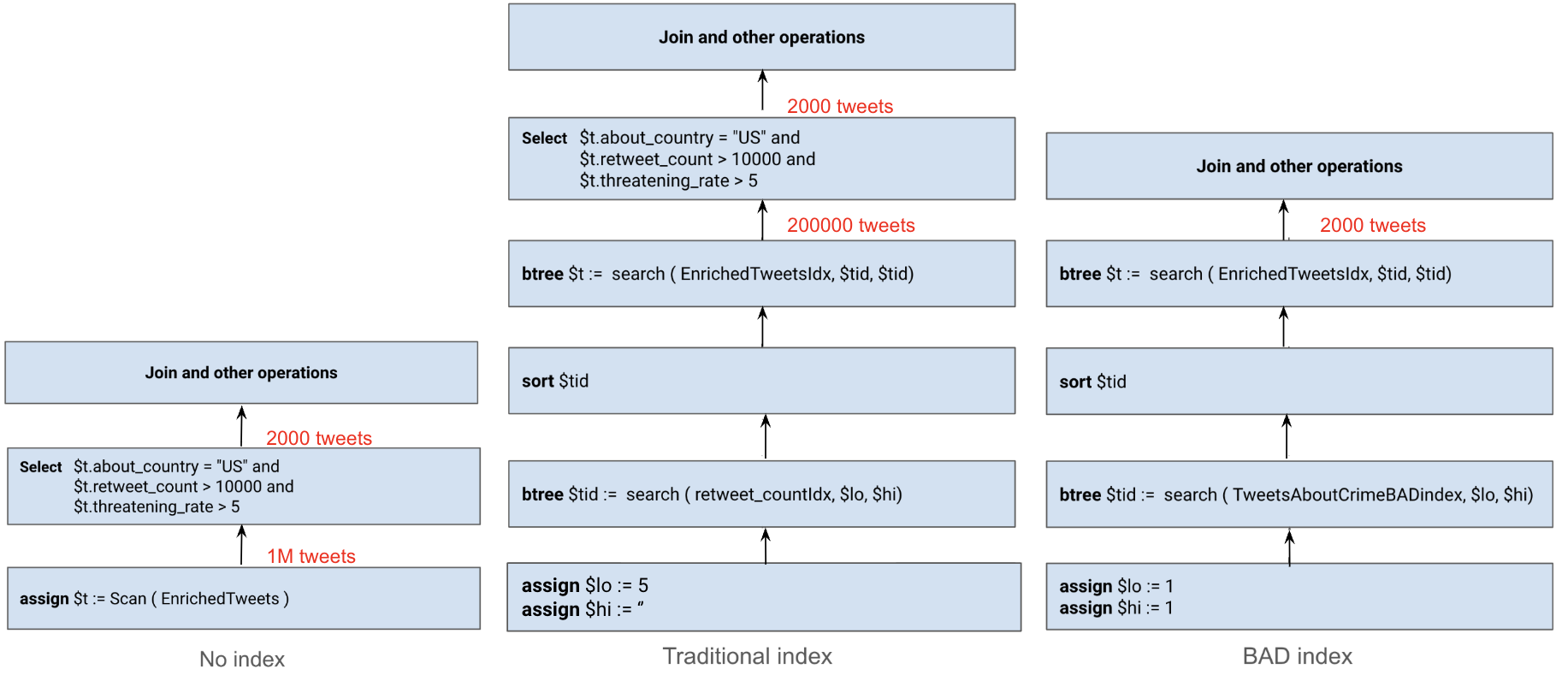}
    \caption{The query plans for retrieving relevant tweets for the {\tt \small TweetsAboutCrime} channel were created without any index, with a traditional index, and with the BAD index.}
    \label{fig:query_plan}
\end{figure*}

To ensure that each channel query effectively utilizes the BAD index, the query plan must be adjusted to replace full dataset scans or the use of other secondary indexes with the BAD index. As explained earlier, regular secondary indexes contain both relevant and irrelevant records, making the BAD index a more efficient choice. Additionally, since the fixed predicates are already processed when data enters the system, they should be removed from the query plan to avoid redundant computations. Figure \ref{fig:query_plan}
 compares the original plan (with and without a secondary index) and the updated plan using the BAD index for the {\tt \small TweetsAboutCrime} channel.
\section{Experimental Evaluation}
We proceed with outlining a series of experiments designed to assess the performance improvements of the optimized BAD platform compared to its original configuration.
We first evaluate the performance of each individual optimization and subsequently examine how the collective implementation of all optimizations provides comprehensive benefits. Lastly, we consider the scale-up and speed-up performance of the fully optimized BAD platform.
It should be highlighted that, previous research has already established the superiority of the BAD platform over a traditional, pieced-together solution with similar functionalities \cite{wang2020subscribing}. Given this, we can assert that an optimized version of the BAD platform would also outperform other comparable systems. Therefore, we have chosen not to repeat those earlier experiments here in this study.

For the following experiments, we used the example application that has been discussed throughout this paper, including its data model, datasets, and  data channels. 
These experiments are conducted on a cluster comprising up to 8 nodes.
Each node has an Intel(R) Xeon(R) CPU E5-2603 v4
@1.70GHz processor, with 64 GB of RAM, 10 TB of HDD, and 2×6-core processors. For all experiments other than the speed-up/scale-up experiments we deployed the BAD platform on a 4-node cluster.

\subsection{Data}
We initialize our experiments by loading the BAD platform with an initial {\tt \small EnrichedTweets} dataset that contains 2 million synthetic tweets. This preloading was performed to ensure that the size of the {\tt \small EnrichedTweets} dataset does not impact the performance of channel execution.
Following the system's initiation, the BAD platform consistently receives 2000 {\tt \small EnrichedTweets} per second, with each {\tt \small EnrichedTweet} being approximately 30 KB in size.
For each channel, we utilized datasets containing 1 million subscribers. Detailed descriptions of these subscriptions will be provided for each experiment.
We have also created a dataset called {\tt \small UserLocations} which is being used in the {\tt \small TweetsAboutCrime} channel and includes people's usernames and their locations; each such record is 38 bytes. We assume that this dataset is continuously updated as the data is received as subscribers change their device's location.
Only the most recent location for each user is maintained in the system.

In this study, synthetic tweets were generated to facilitate a more controlled and rigorous assessment of the system's performance. The use of synthetic data allows for precise manipulation of various field attributes, enabling the evaluation of how different portions of the data contribute to the acceptability for channel and query operations. By systematically altering these parameters, the study provides deeper insights into the system's behavior under varying conditions, offering a robust framework for analyzing the efficiency and scalability of the indexing mechanism. This approach ensures a comprehensive understanding of the impact of data variability on system performance. To demonstrate the broader applicability of our optimizations, we also experimented with real-world datasets, using a publicly available Twitter dataset with continuous updates. Details of this real-world data experiment are discussed in Section \ref{realTweets}.

\subsection{Subscription Aggregation Experiments}
Firstly, we examine the effectiveness of subscription grouping in the {\tt \small TweetsAboutDrugs} channel (Figure \ref{DDL: creating chanel2}). We begin by examining the trade-off presented in Section \ref{Improving Parallelism} regarding the optimal size for subscription groups. We conducted an experiment using a specially created dataset of 1 million subscriptions, all targeting "CA," to evaluate the impact of varying subscription-group sizes on system performance. The experiment was carried out twice: first with a frame size \textit{f}=40 KB, and then with \textit{f}=80 KB. Each original subscription (Figure \ref{fig: query plan} (a)) was approximately 40 bytes in size.

\begin{figure}[!ht]
    \centering
    \includegraphics[width=1\columnwidth]{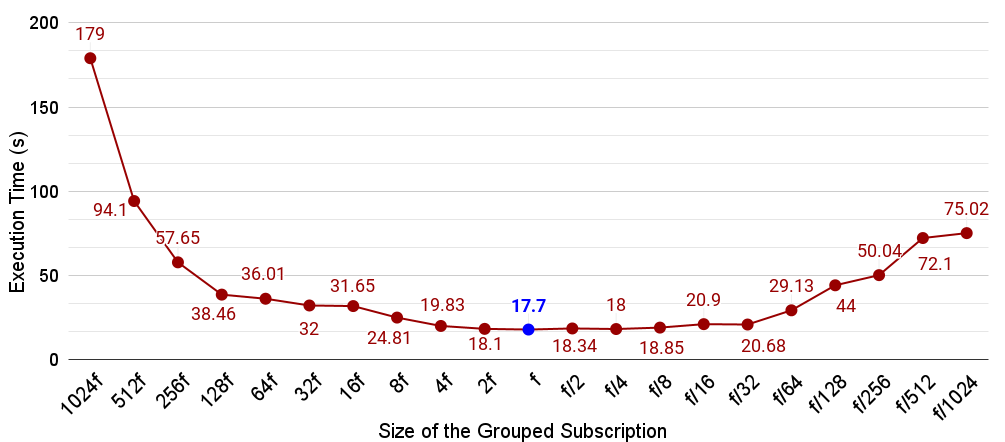}
    \caption{Determining the ideal subgroup subscription size relative to frame size \textit{f}= 40KB. }
    \label{fig: finding the BEST group size}
\end{figure}

For the 40 KB frame size, all "CA" subscriptions were initially consolidated into a single subscription-group, making it 1024 times larger than the frame size. This group was systematically halved to create progressively smaller subgroup sizes. The execution times for these configurations are presented in Figure \ref{fig: finding the BEST group size}.

\begin{figure}[!ht]
    \centering
    \includegraphics[width=1\columnwidth]{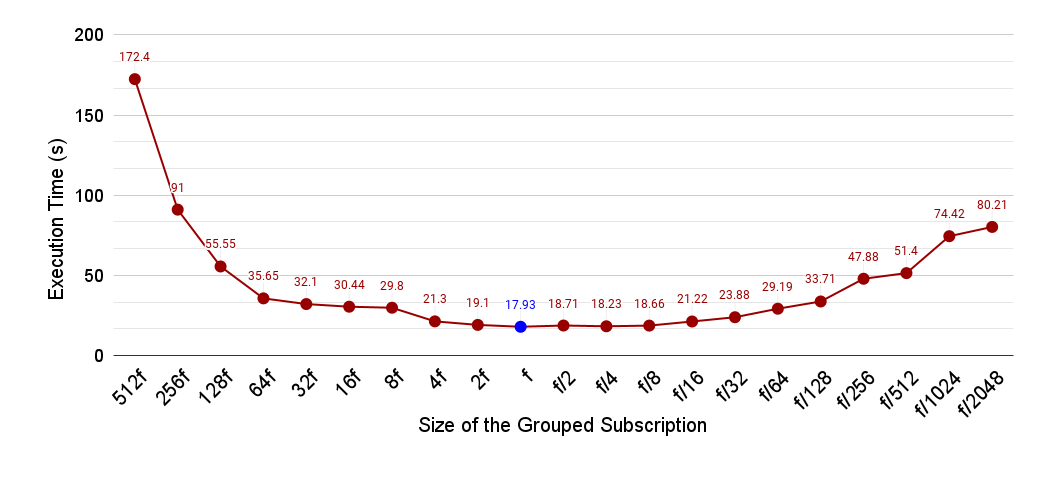}
    \caption{Determining the ideal subgroup subscription size relative to frame size \textit{f} = 80KB. }
    \label{fig: finding the BEST group size for 80KB}
\end{figure}

For the 80 KB frame size, the subscriptions were similarly consolidated into a single subscription-group, this time 512 times larger than the frame size. Again, the group was systematically halved to evaluate the effect of smaller subgroups. The execution times for these configurations are shown in Figure \ref{fig: finding the BEST group size for 80KB}.

Each execution time reflects the duration required to run the channel after processing 10 minutes of incoming tweets at a rate of 2,000 tweets per second. In both figures, the left side (denoted as 1024$f$ for 40 KB and 512$f$ for 80 KB) corresponds to a single subscription-group containing all 1 million subscriptions. The right side ($f$/1024 for 40 KB and $f$/2048 for 80 KB) represents the smallest subgroup configuration, with one subscription per group.

In both figures, the execution time decreases when we have more than a single group, benefiting from enhanced parallelism, but the performance begins to suffer when we create many smaller subgroups, due to increased computational demands. The channel achieves the shortest execution times when the size of each subgroup matches the frame size, effectively balancing computational load and parallelism efficiency. As a result, larger subscription-groups should be split into smaller ones to match the frame size.

Next, we compare the original BAD channel execution time with the optimized BAD that uses aggregated subscriptions (with subgroup sizes of size $f$). We assume the same number of subscriptions, 1 million subscriptions, but now distributed over the 50 US states. In particular, the subscription distribution reflects the real-world population density across U.S. states, ensuring populous states like California and Texas have more subscribers, while less populous states like Wyoming have fewer. For instance, the subscription-group for CA has 118,118 subscriptions which are divided into 116 subgroups, while the subscription-group for Wyoming has just 1,723 subscriptions which are divided into 2 subgroups. Table \ref{tab:aggregating subscriptions} shows the resulting execution times, revealing significant improvements in processing speed when using the grouped subscriptions.
\begin{table}[ht]
\centering
\caption{Channel execution time with and without aggregating Subscriptions}
\label{tab:aggregating subscriptions}
\begin{adjustbox}{width=\columnwidth}
\begin{tabular}{|c|c|c|}
\hline
\textbf{} & \textbf{Original BAD} & \textbf{BAD with Aggregation} \\
\hline
Execution Time (s) & 255.23 & 57.23 \\
\hline
\end{tabular}
\end{adjustbox}
\end{table}


As mentioned earlier, all results generated are relayed to brokers for dissemination to the subscribers. The broker's role primarily involves receiving results, transforming them into a transmittable format, and then disseminating them to users. Table \ref{tab:broker_experiment} illustrates the timing of these operations within a broker tasked with managing results for all "CA" subscriptions. This activity commences upon receiving a drug-related tweet from California by the channel, which is subsequently converted into JSON format \cite{json} and then delivered to its subscribers. The group aggregation offers substantial improvements in broker receiving and processing time. As expected, the time required to then deliver results to the 118,118 CA subscribers does not change between the original and optimized BAD platforms, as the task of distributing the results is the same.

\begin{table}[ht]
\centering
\caption{Timing of different Broker operations in original vs. optimized BAD.}
\label{tab:broker_experiment}
\begin{tabular}{|c|c|c|}
\hline
\textbf{Operation Time(ms)} & \textbf{Original} & \textbf{Optimized} \\
\hline
Receiving Results & 112 & 22 \\
\hline
Converting to JSON & 1109 & 593 \\
\hline
Sending Out & 647 & 632 \\
\hline
\end{tabular}
\end{table}
\subsection{Augmenting The Query Plan Experiments}
\begin{figure}[!ht]
    \centering
    \includegraphics[width=1\columnwidth]{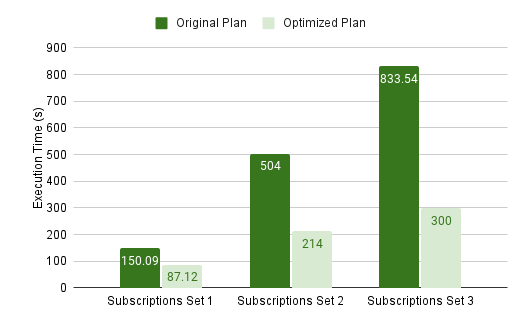}
    \caption{The execution time of channel {\tt\small MostThreatening\\Tweets} with different sets of subscriptions. }

    \label{fig: Augmenting The Query Plan Experiments}
\end{figure}

In this section, we undertake an experiment utilizing the {\tt \small MostThreateningTweets} channel (Figure \ref{DDL: creating chanel}) to demonstrate the impact of augmenting the query plan to integrate the {\tt \small UserParameters} dataset. In particular we focus on varying the proportion of relevant {\tt \small EnrichedTweets} and examine the system's responsiveness and efficiency while adjusting the percentage of {\tt \small EnrichedTweets} that align with subscribers' preferences. We use three subscription datasets that are specifically designed to match a certain percentage of relevant tweets: in set 1 10\% of the subscriptions are matching tweets, in set 2 15\%, and in set 3 20\% of the subscriptions are matching tweets. 
As depicted in Figure \ref{fig: Augmenting The Query Plan Experiments}, re-configuring the query plan to address user interests early on significantly decreases channel execution times for all sets. This optimization is particularly crucial for set 3 due to the higher proportion of relevant data, which necessitates producing a larger number of results. This need for efficiency is much higher when running the original BAD plan. For instance, with the third set of subscriptions, the optimized plan is essential as it allows the channel to process data efficiently and meet time-sensitive deadlines, capabilities that are not as achievable with the original query plan.
Augmenting the query plan with the {\tt \small UserParameters} dataset is particularly beneficial when the incoming data only loosely aligns with users' preferences, 
that is, when processing datasets with lower proportions of relevant tweets, since filtering early reduces unnecessary computation.
When the data closely matches the subscribers' interests, there is no real advantage from filtering (since the datasets are similar), but
the overhead of joining the {\tt \small UserParameters} and {\tt \small EnrichedTweets} is rather negligible because the {\tt \small UserParameters} dataset is very small (containing only a single record per parameter set replicated across the system). 
\subsection{BAD Index Experiments}
To test the BAD index, we employ a variation of the channel {\tt \small TweetsAboutCrime} where we have added more fixed predicates (see Figure \ref{ddl: adding condition}). The goal is to illustrate the impact of the BAD index under different levels of channel selectivity. The full version of the channel contains 5 predicates (marked I through V).
To enhance the selectivity of the channel the predicates are incrementally applied to the channel query. 
The first three predicates (I,II and III) each have selectivity 50\%, while the next two conditions (IV and V) each have selectivity 20\%. By having only the first two conditions we will select around 17\% of the incoming tweets. After applying the first three conditions, the channel selects about 10\% of the incoming tweets. When a fourth condition is added, the selection is reduced to approximately 4.2\%. Finally, applying all five conditions narrows it down significantly to just 0.07\%.

We maintain a dataset called {\tt \small UserLocations}, which includes the saved locations of users. This dataset contains 1 million entries.
Note that changing the subscription dataset does not impact the comparison of execution times between the original BAD system and the optimized version since the BAD index is updated as new tweet records are ingested.
\begin{figure}[!ht]
\footnotesize
\begin{lstlisting}[
           language=SQL,
           basicstyle=\ttfamily,
           showstringspaces=false,
           commentstyle=\color{gray}
        ]
CREATE CONTINUOUS PUSH CHANNEL
TweetsAboutCrime(MyUserName)
PERIOD duration ("PT10M") { 
  SELECT t.text
  FROM UserLocations u, EnrichedTweets t
  WHERE spatial_distance(u.location,t.location)<10
        AND u.username=MyUserName
        AND t.about_country="US"    //(I)
        AND t.retweet_count>10000   //(II) 
        AND t.hate_speech_rate>5    //(III)
        AND t.threatening_rate>5    //(IV) 
        AND t.weapon_Mentioned=true //(V)
        AND is_new(t)};
\end{lstlisting}
\caption{{\tt\small TweetsAboutCrime} channel DDL with additional conditions.}
\label{ddl: adding condition}
\end{figure}

To ensure an equitable comparison, we assess the channel's execution times using a traditional index crafted on the attribute that is most selective under the given conditions. As an example, for the scenario with 2 conditions (I+II), this index is based on the {\tt \small retweet\_count} and when having 4 conditions, the index will be based on the {\tt \small threatening\_rate} as these are the most selective conditions respectively.
 
\begin{figure}[!ht]
    \centering
    \includegraphics[width=1\columnwidth]{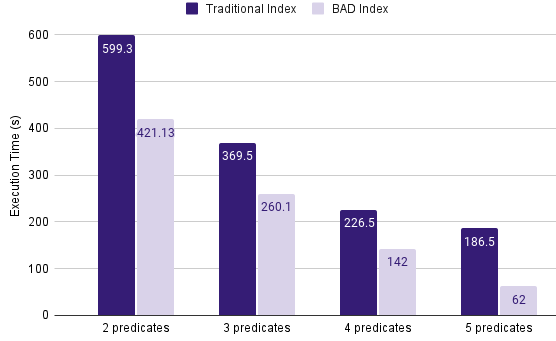}
    \caption{The execution time of the channel {\tt\small TweetsAboutCrime} under varying conditions.}
    \label{fig: BAD index exp}
\end{figure}

 Figure \ref{fig: BAD index exp} shows the channel execution time measured with a traditional index as well as the BAD index.
 The creation of the BAD index significantly enhances the data retrieval capabilities. As illustrated in the figure, it reduces the execution time across various channel selectivities. Notably, the benefits of the BAD index increase as the channels become more selective, enabling more efficient filtering of unsatisfactory records prior to channel execution.
 
\subsection{Comprehensive Performance of Optimized BAD}
We now proceed with an experiment that synthesizes all three distinct optimization strategies that were proposed to enhance the BAD channel performance. 
It is important to note that each optimization may be particularly beneficial for specific scenarios. For instance, the BAD index proves most advantageous in channels which have very selective fixed conditions, the subscription aggregation excels in channels with a limited range of possible parameter values, while customizing the query plan is most effective when the user parameters significantly restrict the dataset that needs to be processed. 

\begin{figure}[!h]
    \centering
    \includegraphics[width=1\columnwidth]{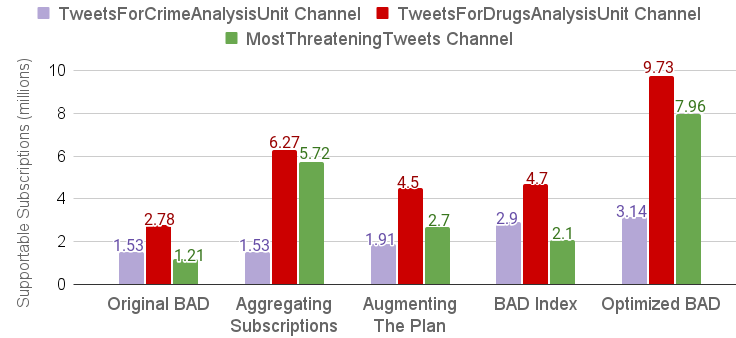}
    \caption{Maximum number of subscriptions supported.}
        \label{fig:max number of subscriptions}
\end{figure}

An important comparison metric is the maximum number of subscriptions that the optimized channel can support; more supported subscriptions reflects the system's improved functionality to process and deliver timely and accurate results to an expanded number of subscribers within set deadlines (in our environment, the maximum number of subscribers that can be supported in the 10 minutes between subsequent channel invocations).
Figure \ref{fig:max number of subscriptions} showcases the enhanced capacity of three specific channels to manage subscriptions using different optimizations; the original BAD, each optimization alone, and the fully optimized BAD as shown. As illustrated in the figure, implementing any combination of the proposed optimizations, or all of them, increases the capacity of each channel to support more subscriptions in the BAD platform.
\subsection{Speed-up and Scale-up Performance} 
\begin{figure}[!h]
    \centering
    \includegraphics[width=0.9\columnwidth]{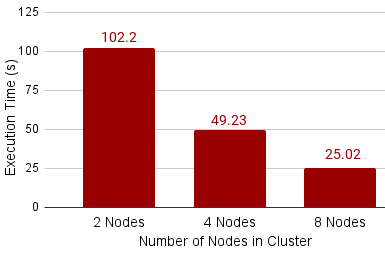}
    \caption{Optimized BAD speed-up performance.}
    \label{fig: speed up experiment}
\end{figure}

\begin{figure}[!h]
    \centering
    \includegraphics[width=1\columnwidth]{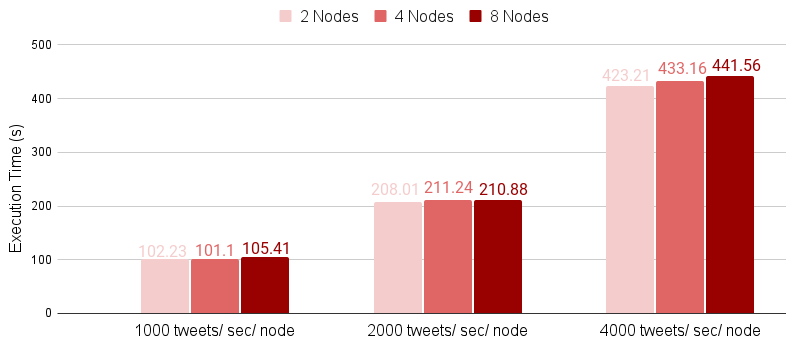}
    \caption{Optimized BAD scale-up performance.}
    \label{fig: scale up experiment}
\end{figure}

For the speed-up and scale-up experiments, we use the channel
{\tt \small TweetsAboutDrugs}. 
To test the speed-up, we increased the cluster size from 2 nodes to 4 and 8 nodes. The rate of incoming tweets was maintained at 2000 per second, with a total of 1 million subscribers. The measured execution times are shown in Figure \ref{fig: speed up experiment}. The optimized BAD platform shows good speed-up performance: as the cluster size doubles, the channel execution time is almost halved.
To measure the scale-up performance we increase the load in proportion to the cluster size (2 nodes, 4 nodes and 8 nodes) by keeping the {\tt \small EnrichedTweet} rate per node constant. We tested with 1 milion users and three different incoming rates per node (1000 EnrichedTweets/sec/node, 2000 EnrichedTweets/sec/node, and 4000 EnrichedTweets/sec/node). For example, in the 1000 EnrichedTweets/sec/node case, the total load for the 2 node cluster was 2000 tweets/sec, and 4000 tweets/sec for the 4 node cluster, etc.
The results appear in Figure \ref{fig: scale up experiment}.
The optimized BAD platform showed good scale-up performance as the channel execution time maintained relative stable as the load and cluster size increase. 

\subsection{Experiments with Real-world Data}
\label{realTweets}
To demonstrate the real-world applicability of our methods, this section utilizes actual tweets collected from Twitter \cite{twitter2024}. According to the study \cite{luu2021persistence}, the most common languages in the overall dataset of real tweets are English, Japanese, Spanish, Arabic, and Portuguese. In the subset of the dataset used for our experiments, English was the dominant language, followed by Portuguese, with the other languages being less represented. As a result, we focused on the two most prevalent languages English and Portuguese. The first channel, called {\tt\small EnglishTrendingTweetsInACountry}, shown in Figure \ref{ddl: real channel}, sends subscribers trending tweets in English (those with {\tt\small retweet\_counts} greater than 100,000). Similarly, the second channel, {\tt\small PortugueseTrendingTweetsInACountry}, targets trending tweets in Portuguese.

Users can subscribe to either channel based on their country of interest, and every 10 minutes, they receive trending tweets from that location. For this experiment, the tweet inflow rate was set to 6,000 tweets per second, which aligns with the average tweet generation rate reported in \cite{xblog2013}. Each tweet averages about 3.5 KB, including details like user data, retweet counts, and location. In total, 1 million subscriptions were generated, with the distribution proportional to each country's population, meaning more populous countries had a larger number of subscribers.

\begin{figure}[!ht]
\footnotesize
\begin{lstlisting}[
           language=SQL,
           basicstyle=\ttfamily,
           showstringspaces=false,
           commentstyle=\color{gray}
        ]
CREATE CONTINUOUS PUSH CHANNEL
EnglishTrendingTweetsInACountry(countryName)
PERIOD duration ("PT10M") { 
  SELECT t.text
  FROM Tweets t
  WHERE t.country=countryName
        AND t.retweet_count>100000
        AND t.lang="en" 
        AND is_new(t)};
\end{lstlisting}
\caption{{\tt\small TrendingTweetsInACountry} channel DDL.}
\label{ddl: real channel}
\end{figure}

 \begin{figure}[!h]
    \centering
    \includegraphics[width=1\columnwidth]{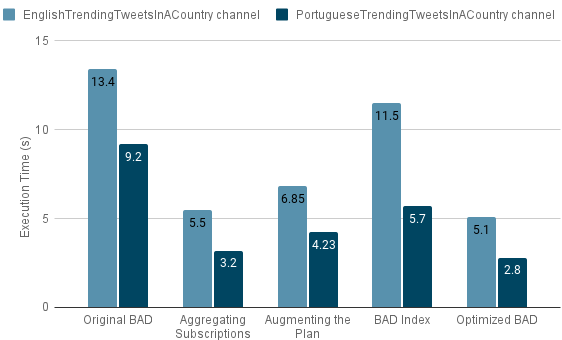}
    \caption{{\tt\small TrendingTweetsInACountry} execution time in different conditions}
    \label{fig: execution time for real channel}
\end{figure}

Figure \ref{fig: execution time for real channel} shows the execution times for both channels. In this figure, we compare the performance of the original BAD system with a traditional index on the most selective field, {\tt \small retweet\_count}, along with various optimizations. As depicted, each optimization reduces execution time for both channels. Notably, the BAD index offers greater time reduction for the {\tt\small PortugueseTrendingTweetsInACountry} channel. This occurs because most tweets are in English, making Portuguese queries more selective and leading to a significant difference in the number of records stored in a traditional index compared to the BAD index.

It is important to note that channels can be created with varying levels of complexity, from simple to highly sophisticated, depending on specific needs. Different channels may benefit from particular optimizations based on their characteristics. However, the experiment in Figure \ref{fig: execution time for real channel} demonstrates that even for simpler channels like {\tt \small EnglishTrendingTweetsInACountry} and {\tt\small PortugueseTrendingTweetsInACountry}, that use non-enriched tweets, the execution time is reduced by 62\% and 70\%, highlighting
the significant and clear benefits of our optimizations.
\section{Conclusions And Future Work}
In this paper, we concentrated on enhancing scalability, performance, and efficiency of a Big Active Data (BAD) platform.
We discussed different example use cases where users use BAD services to monitor a high speed incoming data source like tweets. In order to reduce the execution time and increase the supportable number of users, we introduced three different approaches including: (i) strategically consolidating subscriptions, (ii) revising (augmenting) query plans, and (iii) implementing the BAD index (for early result filtering). Our findings demonstrate a significant enhancement in system performance.
In this paper, our optimization efforts have focused solely on individual channels.
Looking ahead, our future work will focus on strategies for optimizing multiple channels concurrently, which 
holds great potential. For example, we will explore grouping channels and refining the BAD index to synchronize indexing activities across channels.

\balance

\bibliographystyle{abbrv}
\bibliography{sample}
\end{document}